# Operation of a Thick Gas Electron Multiplier (THGEM) in Ar, Xe and Ar-Xe

**R. Alon**[a*]**, J, Miyamoto**[a]**, M. Cortesi**[a]**, A. Breskin**[a]**, R. Chechik**[a]**, I. Carne**[a,b]**,
J. M. Maia**[c,d]**, J.M.F. dos Santos**[d]**, M. Gai**[e,f]**, D. McKinsey**[e] **and V. Dangendorf**[g]

[a] *Department of Particle Physics, Weizmann Institute of Science,*
   *76100, Rehovot, Israel*
[b] *Department of Physics of the University and INFN Milan*
   *I-20133, Milan, Italy*
[c] *Department of Physics, University of Beira Interior,*
   *6201-001 Covilhã, Portugal*
[d] *Instrumentation Center, Department of Physics, University of Coimbra,*
   *3004-516 Coimbra, Portugal*
[e] *Department of Physics, Yale University,*
   *New Haven, CT 06520-8120, USA*
[f] *LNS at Avery Point, University of Connecticut*
   *Groton, CT 06340, USA*
[g] *Physikalisch-Technische Bundesanstalt (PTB)*
   *38116 Braunschweig, Germany*
   *E-mail:* `raz.alon@weizmann.ac.il`

ABSTRACT: We present the results of our recent studies of a Thick Gaseous Electron Multiplier (THGEM)-based detector, operated in Ar, Xe and Ar:Xe (95:5) at various gas pressures. Avalanche-multiplication properties and energy resolution were investigated with soft x-rays for different detector configurations and parameters. Gains above $10^4$ were reached in a double-THGEM detector, at atmospheric pressure, in all gases, in almost all the tested conditions; in Ar:Xe (95:5) similar gains were reached at pressures up to 2 bar. The energy resolution dependence on the gas, pressure, hole geometry and electric fields was studied in detail, yielding in some configurations values below 20% FWHM with 5.9 keV x-rays.

# Contents

## 1. Introduction

The development of novel detectors, having high sensitivity to rare events, with low radioactive background, low energy threshold, and a large mass at a low cost, is crucial for carrying out advanced research in the fields of neutrino, double-beta decay and dark-matter physics [1-7].

There has been a growing interest in utilization of double-phase radiation detectors in these fields [8-12]. In such devices, the incoming particle interacts with a noble liquid, creating ionization electrons which are extracted under electric field into the gas phase and detected after proportional scintillation or gas-avalanche multiplication. In addition, the prompt scintillation of noble liquids can be exploited as well; the scintillation photons may be detected with vacuum photomultipliers in or above the liquid [6], or with gaseous photomultipliers equipped with a photocathode (e.g. CsI [13]). The scintillation photons can be also detected with a photocathode immersed within the noble liquid [14]; here the resulting photoelectrons are extracted from the liquid into the gas phase and detected similarly to the ionization electrons, as described above. Alternately, the ionization electrons and the photoelectrons can also be detected with gaseous photon detectors that record the secondary scintillation light emitted during their transport in the gas phase under high electric fields [15, 16].

In recent years there have been numerous works describing possible solutions to the detection of charges in the gas phase of noble liquids. Some use avalanche multiplication in discrete holes, as to reduce to minimum possible secondary effects due to avalanche-induced photons; others use secondary scintillation, induced by electrons drifting in the gas phase, detected by photomultipliers [6]. In charge-multiplication mode, cascaded Gas Electron Multipliers (GEM), with holes approximately 50 microns in diameter, were shown to operate in noble gases at cryogenic temperatures, including in two-phase conditions [10, 11, 17]; their limited gain could have resulted from condensation of the very cold gas within the tiny holes. There have been other suggestions of using "optimized GEM" multipliers [18], "Large Electron Multipliers" (LEM) [19], MICROMEGAS [20] and more recently Resistive Thick GEMs (RETGEM) [21]. These have millimeter-scale diameter holes drilled in millimeter-scale thick insulator materials.

This work describes the operation properties in noble gases of a "Thick GEM-like Gaseous Electron Multiplier" (THGEM) [22-25].

The THGEM is economically fabricated with a standard Printed-Circuit Board (PCB) technique out of double-clad insulating-material plates; the electrode consists of mechanically drilled holes in the plate, with a chemically etched rim (typically 0.1mm) around each hole; the latter is crucial for reducing the probability of gas breakdowns; higher permissible voltages and hence higher detector gains may then be attained [24]. Due to their simple manufacturing procedure, THGEM electrodes may be fabricated out of a large variety of insulating materials e.g. G-10, Kevlar [23], Cirlex (polyimide) with low natural radioactive background [26], Teflon etc. Due to their proven mechanical robustness THGEM-based detectors can be made very thin and over relatively large areas.

The operation mechanism and properties of the THGEM at atmospheric and at low gas pressure were described in detail in [24, 27]. They can be operated as single-element multipliers or in multi-element (cascaded) mode. An electric potential applied across the THGEM establishes a strong dipole electric field within the holes, responsible for an efficient focusing of ionization electrons into the holes and their multiplication in a gas avalanche process. The resulting avalanche electrons are efficiently extracted from the holes; they can be either collected on an anode or transferred to successive multiplier elements. The THGEM can operate in a large variety of gases; in some of them multiplication factors of $10^5$ with a single THGEM and $10^7$ with a cascaded double-THGEM configuration, were reached at atmospheric pressure [24]. The avalanche process is fast with typical pulse rise-times of a few nanoseconds and rate capabilities in the range of MHz/mm$^2$ [24, 28] at gains of $10^4$; gain stabilities were discussed in [28].

In this article we present the results of our recent studies conducted on the operation of THGEM-based detectors in 1 bar Ar, 0.5-2.9 bar Xe and in 0.1-2 bar of the Penning mixture of Ar:Xe (95:5) [29, 30] at room temperature. Gain and energy resolution were measured for THGEM electrodes of various geometrical parameters in various detector configurations. The experimental research was accompanied by simulation studies.

## 2. The THGEM-based Detector

Measurements were carried out with single-THGEM and double-THGEM detector configurations. The different THGEM electrode geometries employed in this work are summarized in *Table 1*; the electrode's layout is described by its thickness (t), the diameter of the holes (d) and the pitch between holes (a). The holes are always arranged in a hexagonal pattern, and the rim around the holes is always 0.1 mm.

The double-THGEM detector is schematically shown in *Figure 1*; it is composed of a cathode mesh, two Epoxy laminate (FR4) [31] THGEMs in cascade, and an anode mesh; a single-element detector comprised only one THGEM electrode. The components were mounted within a stainless-steel vessel. Measurements in Ar were done under continuous gas flow or in a closed vessel. In Xe and in Ar:Xe (95:5), the chamber and the gas system were pumped down to ~10$^{-5}$ mbar by a turbo-molecular pump, and then filled with gas at different pressures (without baking); in this "closed system", the gas purity was maintained by convection-induced circulation through non-evaporable getters (SAES St 707). The latter, kept at ~200-250 $^{o}$C, were enclosed in a small annex tube connected directly to the chamber. The detector, shown in *Figure 1*, was irradiated with x-rays originating from $^{55}$Fe (5.9 keV) and $^{109}$Cd (22.1 keV) sources. Primary electrons induced by the x-rays in the conversion gap (the space between the cathode mesh and the THGEM) were focused by a drift field E$_{drift}$ into the THGEM holes and multiplied by a single or by two cascaded-THGEM elements. These configurations are denoted single- and double-THGEM. The multiplied charge was further transferred through an induction gap, to a readout anode (e.g. a mesh as shown in *Figure 1*).

| Thickness, t (mm) | Hole Diameter, d (mm) | Pitch, a (mm) | Gas | Detector configuration |
|---|---|---|---|---|
| 0.4 | 0.3 | 0.8 | Argon | Single |
|  | 0.5 | 0.9 |  |  |
|  | 0.6 | 1.2 |  |  |
|  | 0.8 | 1.3 |  |  |
|  | 0.3 | 1.0 | Xenon | Single/ Double |
|  | 0.5 | 0.9 | Ar:Xe (95:5) | Single/ Double |
| 0.8 | 0.4 | 0.9 | Argon | Single |
|  | 0.6 | 1.2 |  |  |
|  | 0.8 | 1.3 |  |  |
|  | 0.4 | 1.2 | Xenon | Single/ Double |

*Table 1: THGEM geometries, detector configurations and gases investigated in this work*

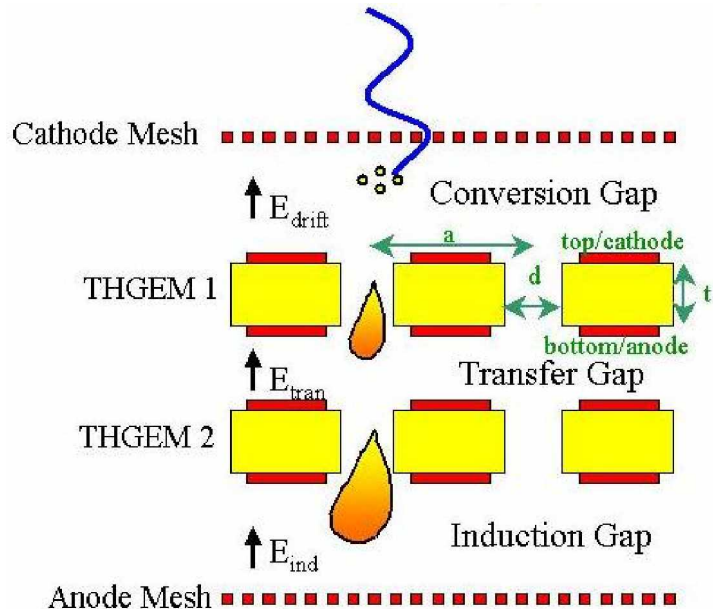

*Figure 1: Schematic Drawing of the Double-THGEM Detector*

All THGEM electrodes were biased with independent high-voltage power supplies (CAEN N471A) through 20 MΩ resistors and the signals were read through decoupling capacitors. Charge pulses were recorded from the last amplification stage. Signals were processed with charge-sensitive preamplifiers, further amplified and shaped with linear amplifiers; the pulse-height spectra were analyzed with a multi-channel analyzer (MCA). The whole chain of electronics from the preamplifier to the MCA was calibrated to measure absolute avalanche charges.

## 3. Results

We present results regarding electron multiplication, gain and energy resolution, with single- and double-THGEMs, operating in 1 bar Ar, 0.1-2 bar Ar:Xe (95:5) and in 0.5-2.9 bar Xe.

### 3.1 Gain

*Figure 2* shows typical gain curves measured in 1 bar Ar, Ar:Xe (95:5) and Xe with double-THGEM detectors having thicknesses of 0.4 mm and hole diameters of 0.3, 0.4 and 0.5 mm, in a closed system. The maximum effective charge gains correspond to the appearance of discharges or spontaneous electron emission. In Xe and Ar:Xe (95:5), the maximum gains reached with two cascaded THGEMs, and with optimized drift-, transfer- and induction-fields [24], were above $10^4$ at atmospheric pressure. Ar measurements in a closed system (*Figure 2*) yielded lower gain than those taken in gas-flow mode (see *Figure 3* and discussion below).

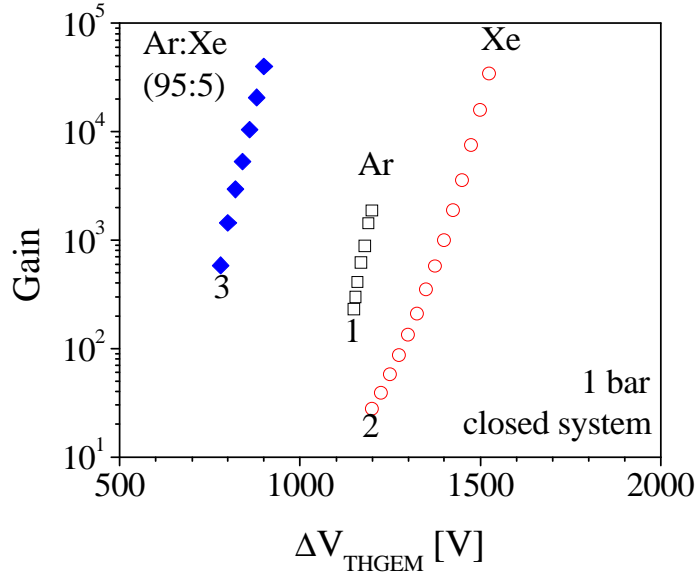

*Figure 2: Gain curves of a double-THGEM operated in a closed system, with internal gas circulation through a getter after evacuation to high vacuum: (1) Ar: t=0.4mm, d=0.5mm, a=0.9mm; (2) Xe: t=0.4mm, d=0.3mm, a=1mm; (3) Ar:Xe (95:5): t=0.4mm, d=0.5mm, a=0.9mm.*

Unless otherwise mentioned, all the following measurements in Ar were done in a gas-flow mode.

Different THGEM-electrode geometries were investigated in 1 bar Ar, as shown in *Figure 3*. In terms of gain, they all provided rather similar results, except for one, in which the hole diameter was twice as large as the electrode's thickness (curve 5 in *Figure 3*). As pointed out in [24], maximal gain is typically reached when the ratio t/d ~1; in this respect, curve 5 should be compared to curve 8 in *Figure 3*, which was measured with same hole diameter and pitch but with a double thickness.

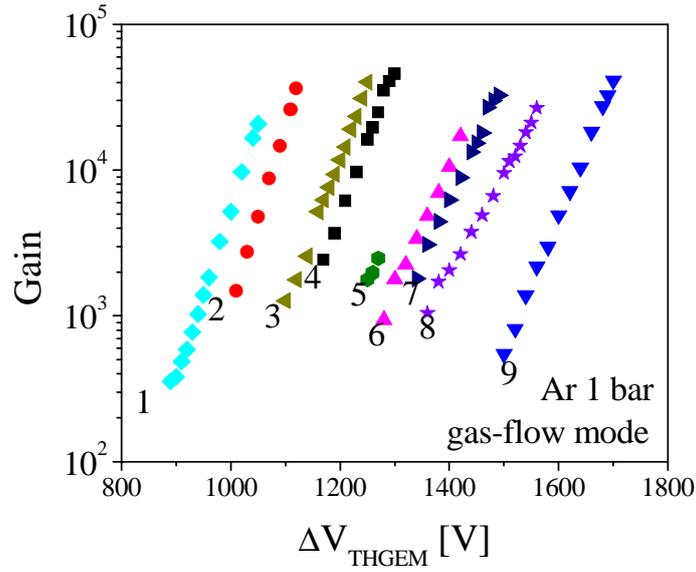

*Figure 3:* Double-THGEM gain curves measured with 5.9keV x-rays in various geometries in Ar, in a gas-flow mode at 1 bar: (1) t=0.4mm, d=0.3mm, a=0.7mm; (2) t=0.4mm, d=0.3mm, a=0.8mm; (3) t=0.4mm, d=0.5mm, a=0.9mm; (4) t=0.4mm, d=0.6mm, a=1.2mm; (5) t=0.4mm, d=0.8mm, a=1.3mm; (6) t=0.8mm, d=0.4mm, a=0.9mm; (7) t=0.8mm, d=0.6mm, a=1.2mm; (8) t=0.8mm, d=0.8mm, a=1.3mm; (9) t=0.8mm, d=0.6mm, a=1mm

In the Xe measurements, the electric fields in the different gaps were raised with proportion to the pressure in order to maintain constant reduced electric field (E/p) values. Within the THGEM's holes, the E/p values were limited by the maximum voltage the THGEM could hold. This resulted in a continuous decrease in the maximum reachable gain with increasing pressure. The results for single- and double-THGEMs with electrode thicknesses of 0.4 and 0.8 mm are shown in *Figure 4*.

*Figure 5* shows gain curves measured with single- and double-THGEMs in the Ar:Xe (95:5) mixture at 1 bar, in a closed vessel after evacuation to high vacuum. A gain curve in 1 bar Ar, measured with the same detector in similar conditions, is shown for comparison.

*Figure 6* shows gain curves measured with a double-THGEM detector in Ar:Xe (95:5) in a pressure range of 0.1-2 bar. The measurements were done in a closed vessel, after evacuating to high vacuum.

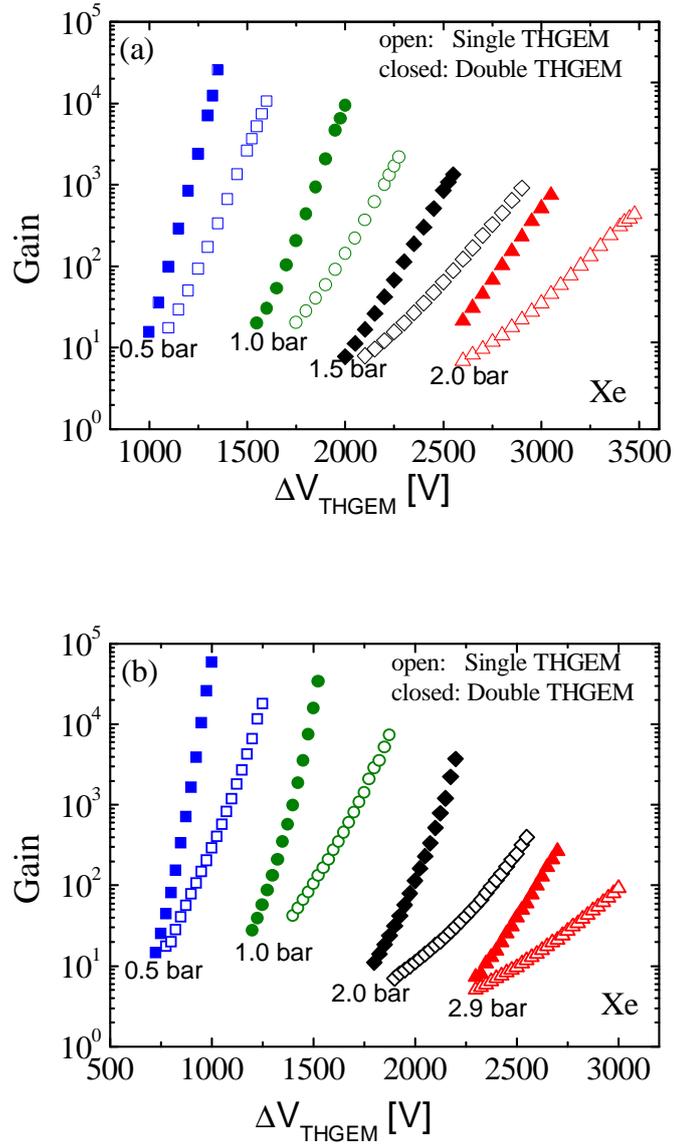

*Figure 4: Gain curves in single- and double-THGEM operated in Xe at various pressures: (a) t=0.8mm, d=0.4mm, a=1.2mm; (b) t=0.4mm, d=0.3mm, a=1mm. Measurements at pressures of 0.5 and 1 bar were done with 5.9 keV x-rays and with 22.1 keV x-rays at pressures above 1 bar.*

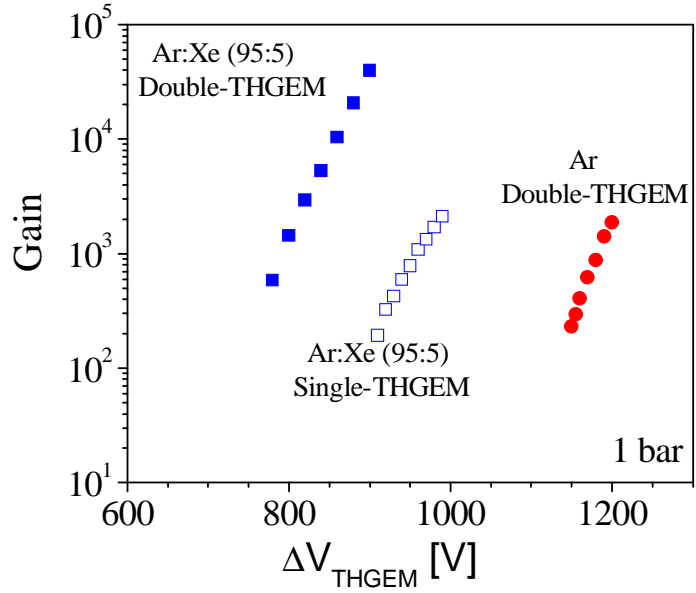

*Figure 5:* *Gain curves in single- and double-THGEM operated in Ar:Xe (95:5) at 1 bar, in a closed vessel with internal gas circulation through a getter. The gain curve in Ar, measured with the same THGEM electrodes and in the same conditions is shown for comparison. Detector geometry: t=0.4mm, d=0.5mm, a=0.9mm.*

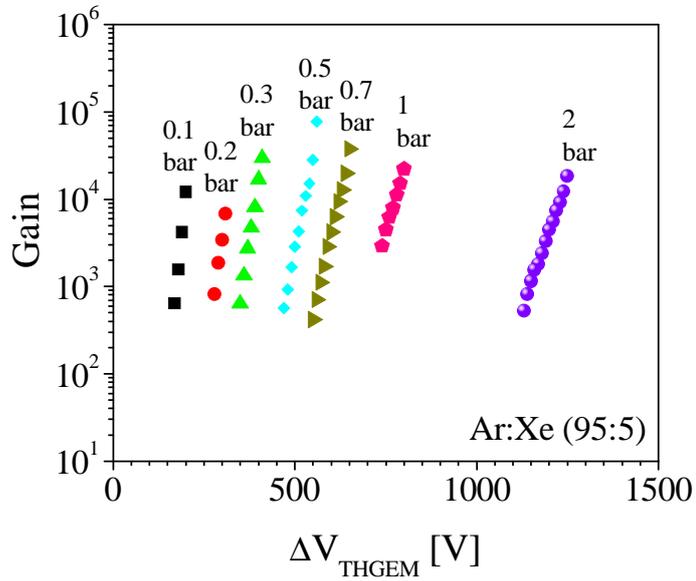

*Figure 6:* *Gain curves taken in Ar:Xe (95:5) with a double-THGEM detector in different pressures, as indicated in the figure. Detector geometry: t=0.4mm, d=0.5mm, a=0.9mm.*

## 3.2 Energy Resolution

Pulse-height spectra recorded in Ar and Xe in different THGEM configurations, with $^{55}$Fe and $^{109}$Cd x-rays are shown in *Figure 7*. *Figure 8* shows pulse-height spectra recorded in Ar:Xe (95:5) with $^{55}$Fe 5.9 keV x-rays.

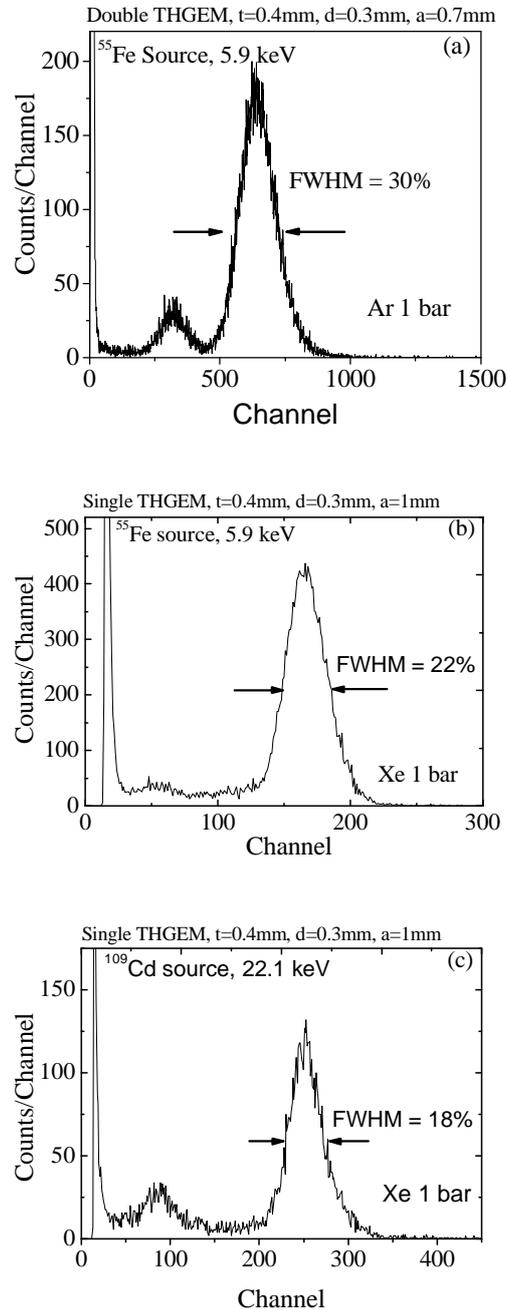

***Figure 7:*** *Pulse-height spectra recorded in single- and double-THGEM detectors of geometries indicated in the figures, at 1 bar in: Ar (a) and Xe (b), with $^{55}$Fe 5.9 keV x-rays and Xe (c) with $^{109}$Cd 22.1keV x-rays, for detector gains ~$10^3$ and ~$10^4$ in single- and double-THGEM respectively.*

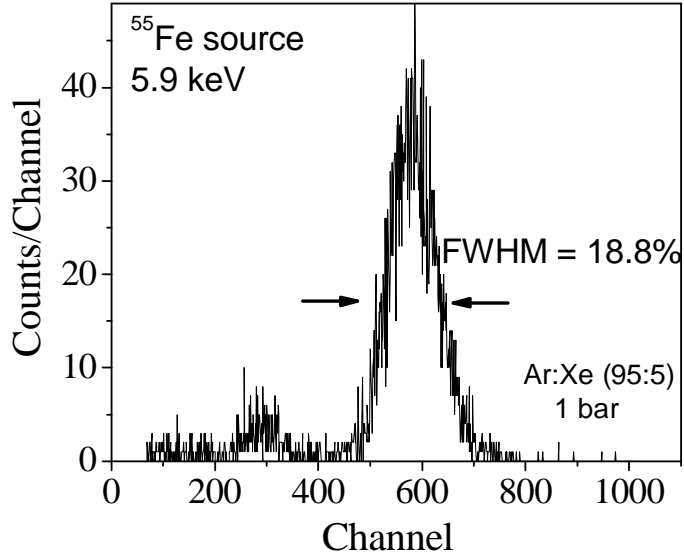

*Figure 8: Pulse-height spectra recorded with a double- THGEM detector with geometry t=0.4mm, d=0.5mm, a=0.9mm, at 1 bar Ar:Xe (95:5) with $^{55}$Fe 5.9 keV x-rays. Detector gain ~$10^4$.*

The energy resolution is known to depend on the drift field, which defines the electron diffusion. With hole multipliers, the resolution is first and foremost defined by the ratio of drift-to-hole fields, since this field ratio defines the electron transfer efficiency, namely the efficiency to bring a single electron from the drift gap into the multiplication region inside the hole. This ratio should not be too large or the electron will be collected at the metal top (cathode) face of the THGEM. The dependence on the drift field is shown in *Figure 9* for Xe with 5.9 keV and 22.1 keV x-rays, and in *Figure 10* for Ar with 5.9 keV x-rays. At fixed gain (hole field), the resolution indeed deteriorates, and more pronouncedly so with the smaller hole diameter.

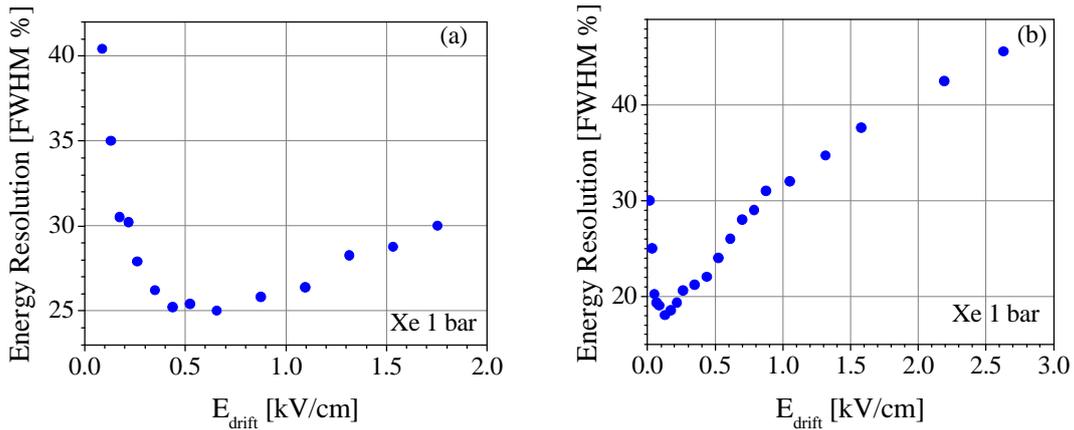

*Figure 9: Energy resolution of a single-THGEM versus drift field, measured in 1 bar Xe: (a) 5.9 keV x-rays, t=0.8mm, d=0.4mm, a=1.2mm; (b) 22.1 keV x-rays, t=0.4mm, d=0.3mm, a=1mm. Detector gain ~$10^3$*

In full agreement with the above mentioned electron transfer efficiency dependence on the drift-to-hole field ratio, an improvement of the energy resolution with gain increase, up to gain values of the order of ~$10^4$ was measured in Ar (*Figure 11*); this occurred for various values of the drift field.

The curves of the energy resolution versus gain in Ar:Xe (95:5) have similar shapes to those measured in Ar; however, with better energy resolution, as shown in *Figure 12*.

*Figure 13* shows the best energy resolution obtained in Ar:Xe (95:5) over a pressure range of 0.4 to 2 bar.

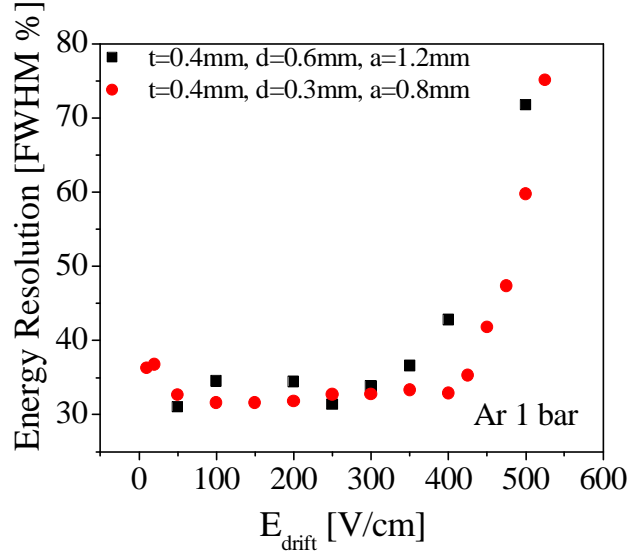

***Figure 10:*** *Energy resolution versus drift field in a double-THGEM operated with 5.9 keV x-rays in 1 bar Ar. Detector gain ~$10^4$*

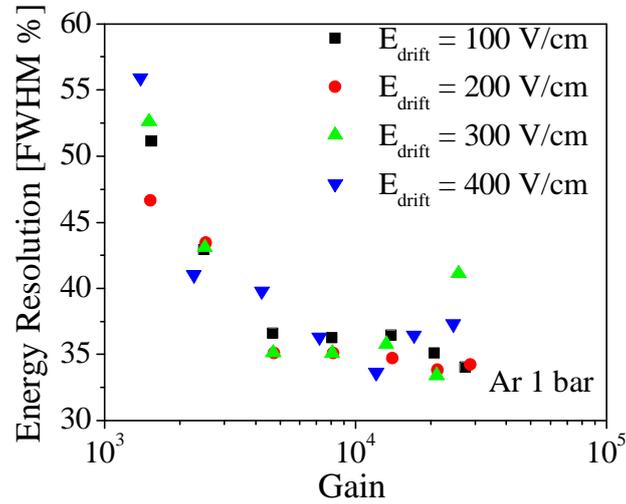

***Figure 11:*** *Energy resolution versus gain in a double-THGEM detector operated in a gas-flow mode in 1 bar Ar, at different drift fields. Detector geometry: t=0.4mm, d=0.3mm, a=0.8mm.*

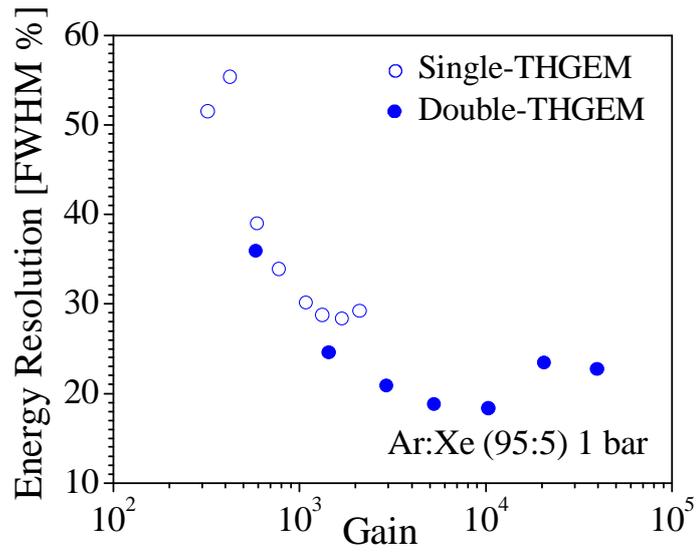

*Figure 12:* *Energy resolution versus gain in single- and double-THGEM detector operated in 1 bar Ar:Xe (95:5). Detector geometry: t=0.4mm, d=0.5mm, a=0.9mm.*

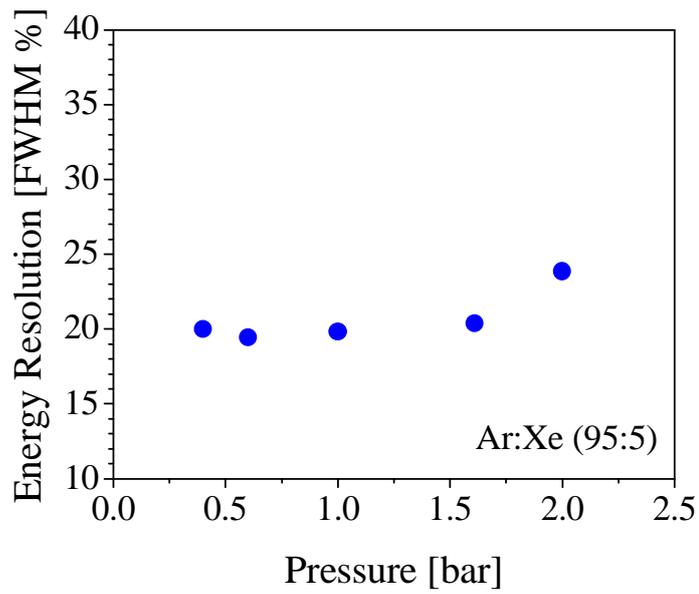

*Figure 13:* *Energy resolution versus pressure in a double-THGEM detector in Ar:Xe (95:5). Detector geometry: t=0.4mm, d=0.5mm, a=0.9mm. Drift field is 100 V/cm. Detector gain ~$10^4$.*

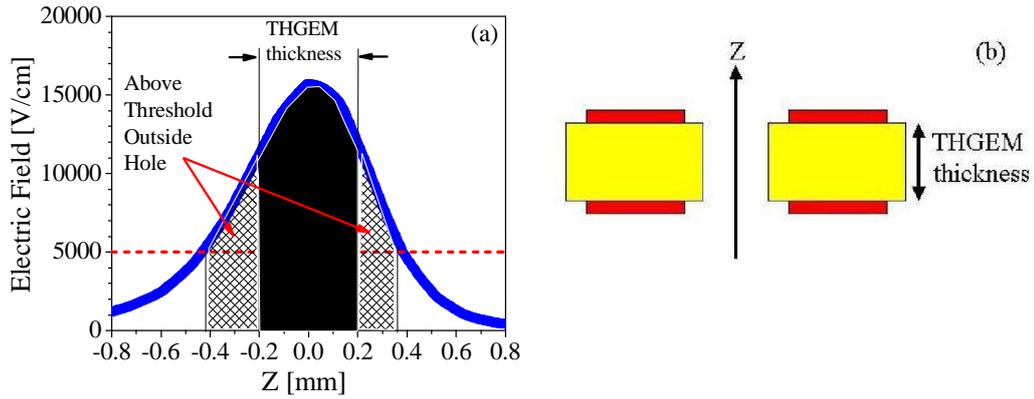

*Figure 14: Maxwell [32] simulation results (a) of the electric field inside the THGEM hole along its Z axis (shown in b), at 1 bar. The chosen multiplication threshold is indicated by a dashed line, and the multiplication regions are indicated by the dark and striped areas, inside and outside the hole, respectively. The field penetration ratio is the striped area divided by the sum of the striped and the dark areas.*

### 3.3 Simulations

Aside from electron transfer efficiency, the energy resolution depends on the multiplication mechanism, and specifically on its uniformity. Due to the dipole nature of the hole field, there is always field penetration [24] from the hole into the gaps above and below the THGEM, which depends on the hole geometry and the field strength across the hole and in the gaps. This field penetration can affect the energy resolution. To understand the dependence of the energy resolution on the geometry and the electric field we carried out a simulation study of the electric field magnitude inside and outside the hole, using Maxwell 3D [32]; the study was done for different hole diameters, keeping a fixed gain ($10^4$); the maximum field value inside the hole was about the same for all geometries, but its shape differed. A multiplication threshold was set to electric fields above 5 kV/cm, corresponding to an approximate onset of charge multiplication in the investigated gases [33-34]. As an estimate of the field penetration outside the hole, the integrated area under the electric field curve was used, as seen in *Figure 14* (a). The field penetration ratio, i.e. the ratio between the area under the curve where the field is above the threshold and outside the hole (striped areas in *Figure 14* a), to the area under the curve where the field is above the threshold (sum of striped and dark areas in *Figure 14* a) was calculated.

*Figure 15* shows the field penetration ratio versus the hole diameter for three different THGEM thicknesses. The 0.6mm thick electrode was not used in any of the experiments; its simulation results are shown for comparison.

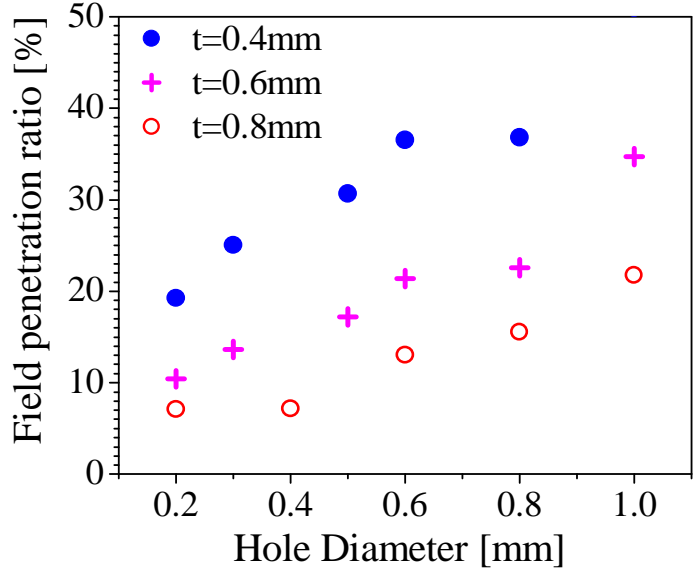

*Figure 15: Ratio of field outside the hole and above threshold, to the field above threshold, versus hole diameter for different THGEM thicknesses.*

## 4. Summary and discussion

Measurements in this work were done at room temperature, with single- and double-THGEM detectors, with electrodes having various geometric parameters presented in *Table 1*. Gains above $10^4$ were reached practically in all geometries, in all gases investigated at 1 bar: Ar, Xe and Ar:Xe (95:5). The dependence of the maximum-gain in 1 bar Ar on the hole diameter, measured with 5.9 keV x-rays, (*Figure 16*) indicates that the maximum gain was achieved with holes smaller than 0.6mm in diameter. The gain limit for the larger-diameter holes could be explained by larger electric-field penetration from the hole into the surrounding gaps, shown in *Figure 17*; the latter requires operation at higher voltages for reaching similar gains; it also causes avalanche extension outside the holes, accompanied by photon emission – inducing secondary effects. Measurements in Xe were performed over a pressure range of 0.5-2.9 bar; an increasing drop in the maximum achievable gain was observed at pressures >2 bar, as shown in *Figure 4*. In Ar:Xe (95:5), gains above $10^4$ were measured practically over the entire pressure range 0.1 – 2 bar (*Figure 6*). The higher gain at 2 bar compared to that reached in Xe could be explained by the significantly lower operation voltages in this Penning mixture.

It should be noted that in some cases (particularly in Ar as seen in curve 1 in *Figure 2*) the detector's gain limit was lower after evacuating the detector vessel to high vacuum conditions, prior to gas introduction. This is attributed to the absence of impurities, which often act as avalanche-photon "quenchers". The effect of the quencher is more beneficial in Ar, which emits more energetic avalanche photons compared to Xe (at wavelengths of ~120 nm for Ar compared to ~170 nm for Xe); these photons induce secondary effects (e.g. photoelectron emission from electrodes and walls) which limit the operation stability at higher gain. Lower gain limits at high gas purity could also result, to some extent, from charging up of the electrode's FR4 substrate which after extended pumping (e.g. of water molecules) may have an increased surface resistivity. These effects are the subject of current studies. After evacuation followed by gas filling and circulation through getters, 20-fold higher gains were reached with the Penning mixture [29, 30] of Ar:Xe (95:5) compared to those of pure Ar.

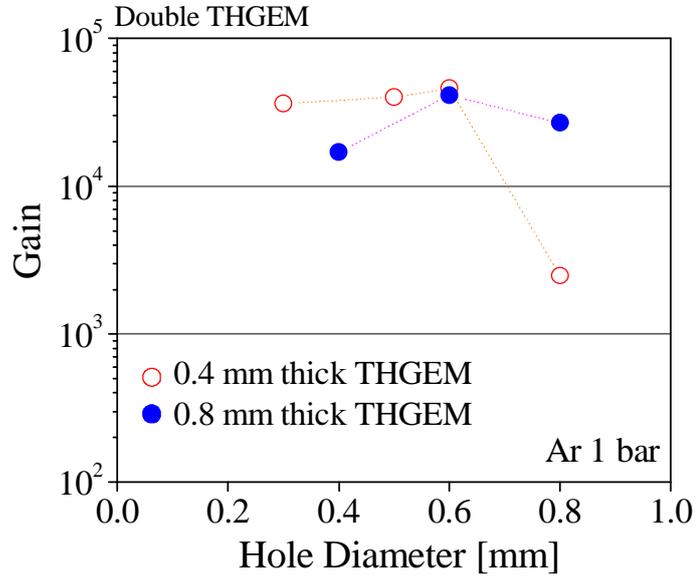

*Figure 16: Maximum gain versus hole diameter measured with double-THGEM in 1 bar Ar for 0.4 and 0.8 thick THGEM plates*

At cryogenic temperatures, the maximum achievable gain in all gases is expected to drop due to the increase in gas density, as recently demonstrated with cascaded GEMs [10, 11] and in Resistive Electrode Thick GEMs (RETGEM) [21].

Best energy resolutions reached in Ar with 5.9 keV x-rays using double-THGEM were of the order of 30% FWHM. These results are similar to those recently measured with RETGEM in Ar at similar working conditions [35]. The resolution in Xe for 5.9 keV x-rays using single-THGEM reached values of 21%-22% FWHM over the pressure range of 0.5-1 bar, respectively, and 27% using double-THGEM. In Ar:Xe (95:5) Penning mixture a better energy resolution was measured, of ~20% FWHM in a double-THGEM over the pressure range of 0.4 – 2 bar with a slight increase to ~24% at 2 bar (*Figure 13*).

The difference between the energy resolutions measured in the different gases originates from the differences in the statistical fluctuations in the numbers of primary and avalanche electrons; these are function of the W-values (the average energy per an electron-ion pair) [36], Fano factors [34, 36] and the parameters characterizing avalanche statistics [37, 38]. The high gains reached in Ar:Xe(95:5) at low multiplication fields and the superior energy resolutions result from the lower W-value of 23.2 eV [36], and wave-length shifting of the high energy avalanche photons of Ar to Xe wavelengths.

The best energy resolutions in 1 bar Ar were achieved with holes smaller than 0.6 mm in diameter. THGEM plates of thickness 0.4 mm with 1 mm diameter holes (results not shown) yielded energy resolutions of 46% and lower gains. The electric field penetration from larger-diameter holes into the surrounding gaps causes gain fluctuations due to partial amplification outside the hole, affecting the energy resolution.

It was shown earlier, in other multiplication geometries [39], that the avalanche fluctuations are reduced at smaller avalanche-formation volumes. This supports the better energy resolution observed with smaller-diameter holes.

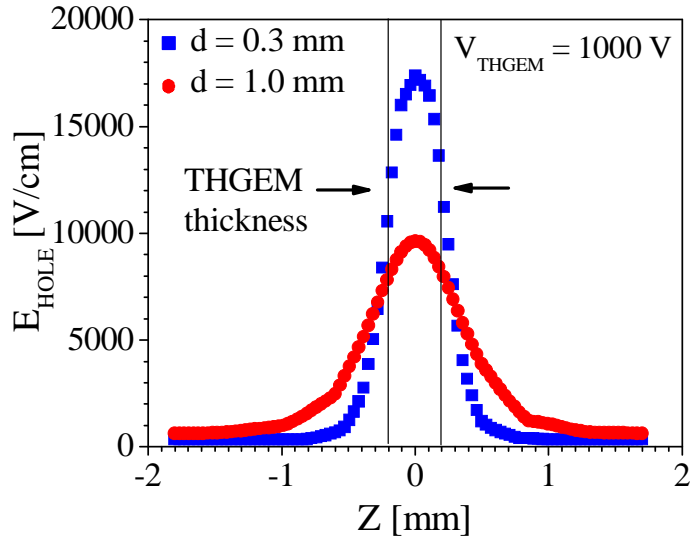

*Figure 17: Maxwell [32] simulation of the electric field inside the THGEM hole along the Z axis for two different hole diameters. THGEM thickness: 0.4mm; potential across the hole: 1000V*

*Figure 15* clearly indicates that for a constant gain, for each THGEM plate thickness, the field penetration ratio increases with hole diameter. For a constant hole diameter, the field penetration decreases with the THGEM plate thickness. This may explain the difference in the slopes of the energy resolution versus drift-field plots shown in *Figure 9* (a) and (b). Since the field penetration is less significant with the 0.8 mm thick electrode (for equal hole diameter), the energy resolution is less dependent on the drift field, compared to that of the 0.4 mm thick electrode.

The energy resolution dependence on the drift field has a clear minimum, both in Ar and in Xe (*Figure 9* and *Figure 10*). A drift-field correlated behavior of the charge collection into the holes, affecting energy resolution, was also observed in GEM detectors [40]. At low drift fields primary electrons are lost due to diffusion and recombination. Transverse diffusion coefficient in Ar has a minimum value at electric fields around 200 V/cm which matches the experimental energy resolution results [41]. At higher field values electrons can be lost due to their collection on the first THGEM top (cathode) electrode instead of entering into the holes. *Figure 9*(a) and *Figure 10* show constant values of the energy resolution over a broad range of the drift field. The resolution changes more drastically in *Figure 9*(b) with the thinner THGEM. This may also be attributed to the stronger field penetration in the case of thinner THGEM plates, as shown in *Figure 15*.

Although higher gains and improved energy resolution were demonstrated in this work with the Ar:Xe(95:5) Penning mixture, it is probably not usable in the gas phase of a two-phase detector; it would cause unnecessary elevated pressure of Ar in a LXe-based detector. A possible LAr detector with dissolved Xe from a Ar:Xe mixture, indeed would not require Xe concentration of 5% as used in this work. Even with much lower Xe additives, the Penning effect exists and minute fractions of dissolved Xe in LAr will shift the photon spectrum into that of Xe.

In the case of the LXe detector a possible solution would also be other Penning mixtures based on Xe as parent gas. The results of such studies in Xe:$CH_4$ were reported in [10].

## 5. Outlook

The data presented above relate to THGEMs in noble gases, at room temperature. The results permit a more extensive study planned at cryogenic temperatures for evaluating the operation properties of THGEM electrodes in two-phase detectors. The latter, recently successfully operated with cascaded GEMs in Ar and Xe [10], could be good candidates for ionization and scintillation signals recording in large-volume Dark-Matter, double-beta decay and neutrino experiments, and in gamma-cameras for PET. THGEMs are expected to offer more stable operation in cryogenic conditions due to lower condensation effects in the ten-fold larger holes compared to GEMs. THGEM electrodes of low-radioactivity materials, e.g. Cirlex [26], have been investigated for rare-event experiments; they are expected to yield a more economic solution for large-volume detectors and significantly lower radioactive background compared to photomultiplier tubes. A R&D project of a liquid-xenon Gamma Camera for medical imaging, incorporating THGEM photon detectors, is also in course in cooperation with Subatech – Nantes.

### Acknowledgements

This work was partly supported by the Israel Science Foundation, grant No 402/05, by the Weizmann-Yale collaboration program of the American Committee of Weizmann Institute of Science (ACWIS), New York, by a grant No 3-3418 of the Israel Ministry of Science, Culture & Sport within a France-Israel Scientific Cooperation and by the Foundation for Science and Technology (FCT), Portugal, through the project POCI/FP/81980/2007. Scientific discussions with Dr. D. Thers of Subatech – Nantes, technical support by Mr. M. Klin of the Weizmann Institute and efficient cooperation of Print Electronics [31] are greatly acknowledged. A. Breskin is the W.P. Reuther Professor of Research in The Peaceful Use of Atomic Energy.